# A review of quality frameworks in information systems

Thanh Thoa Pham Thi, Markus Helfert

School of Computing Dublin City
University, Ireland
Thoa.Pham@computing.dcu.ie
Markus.Helfert@computing.dcu.ie

**Abstract:** Quality is a multidimensional concept that has different meanings in different contexts and perspectives. In the domain of Information system, quality is often understood as the result of an IS development process and as the quality of an IS product. Many models and frameworks have been proposed for evaluating IS quality. However, as yet there is not a commonly accepted framework or standard of IS quality. Typically, researchers propose a set of characteristics, so-called quality factors contributing to the quality of IS. Different stakeholders' perspectives are resulting in multiple definitions of quality factors of IS. For instance, some approaches are based on the IS delivery process for the selection of quality factors; while some other approaches do not clearly explain the rationale of their selection. Moreover, often relations or impacts among selected quality factors are not taken into account. Quality aspects of information are frequently considered isolated from IS quality. The impact of IS quality on information quality seems to be neglected in most approaches. Our research aims to incorporate these levels, by which we propose an IS quality framework based on IS architecture. Considering user and IS developer's perspectives, different quality factors are identified for various abstraction levels. Besides, the presentation on impacts among different quality factors helps to retrieve the root cause of IS defects. Thus, our framework provides a systematic view on quality of information and IS.

# 1 Introduction

Quality is a multidimensional concept which has different meanings in various contexts and perspectives. In the literature, among several popular research approaches, there are basically two axes which found a large interest:

- Data quality [BP95; PLW02; WS96] which refers to the quality of data in database and data warehouse.
- Information system quality [DR06; PSV94] and software quality [Wo06; VWW93; As96] which are related to the quality of IS products and software products.

Due to the heterogeneous perspectives of multiple IS stakeholders, researches on the IS quality have a diversity on definitions of criteria/factors [DR06]. For instance, in some approaches, the selection of quality factors is based on key factors of the IS delivery process [DR06], while with other approaches the selection of quality factors and criteria seem to be made arbitrarily" [KP96]. This shows that there is no common rationale of criteria selection. Finally, so far to our best knowledge, there is no commonly accepted model or framework on IS quality.

Besides, most research approaches on data quality, typically study data quality at the database and data warehouse level, isolated from the IS quality. However data are stored in databases according to a pre-defined structure and rules specified during the IS development process. Therefore, data quality is also impacted by specifications and quality of specifications in the IS development process.

In the IS context, the user and IS Developer are the principal stakeholders. Besides, in our point of view, studying IS quality cannot be isolated from the structure of IS or IS architecture.

In this paper, we propose an IS quality framework based on IS architecture. This takes into account the perspectives of user and IS Developer. IS architecture is a generic and logical model/framework that describes IS structures including its elements and relation among them. Any IS is characterized by this structure. Following the approach for IS architecture, we introduce a set of quality factors that contributes to IS quality. We present impacts and relations among quality factors including impacts on data quality. Our quality framework allows to contribute to manage the quality in IS and retrieve root causes of defects.

The paper is organized as follows. Section 2 concerns background in information system architecture that we base on; we also present, in this section, software quality and information system quality concepts. Section 3 introduces our quality framework and discussions on it. Our conclusion is presented in Section 4.

## 2 Background

In this section, we review different approaches to information systems architectures, information systems quality and software quality.

### 2.1 Information Systems Architecture

IS architecture presents an integrated structural design of a system, its elements and their relationships depending on given system requirements [BS98]. IS architecture is also defined as "a logical structure for classifying and organizing the descriptive representations of an Enterprise that are significant to the management of the Enterprise as well as to the development of the Enterprise's Systems" [Za87]. Mostly an IS is considered under different aspects such as information, function, organization, resource and different level abstractions that depend on the IS stakeholders' perspectives.

For instance, Zachman [Za87] defines IS architecture as a matrix in which the *rows* represent different *abstraction levels*, each of which is said to represent the perspective of a named role in the organisation such as Scope (contextual level), Owner (conceptual level), Logical Designer (logical level), Builder (Physical level); the *columns* represent different *views/ aspects* of IS such as Data, Function, Network, Organization, Time and Motivation. Each cell of this matrix describes relevant models or diagrams used for depicting the correspondent view in the correspondent abstraction level.

Another framework for the presentation of IS and for the modelling of the entire enterprise is the GERA modelling framework [BS98]. The GERA framework describes what models of the enterprise are needed and maintained during the enterprise life.

This framework structures an IS at three dimensions for defining the scope and content of enterprise modelling such as

- *Life-cycle Dimension* that provides the controlled modelling process of enterprise entities according to the life-cycle activities. This dimension is similar to different abstract levels in relation to different IS stakeholder's perspectives in the Zachman framework.
- *Genericity Dimension* that provides the controlled instantiation process from generic and partial to particular.
- *View Dimension* that provides the controlled visualisation of specific views of the enterprise entity. Some among them are information, function, organization and resource views.

Each cell in the framework presents models of a correspondent view, a correspondent stage in the life-cycle and an appropriate degree of generality. The IS architectures helps to understand the IS structure, decrease the complexity of the IS development through different views and levels and it is independent with IS development process.

## 2.2 Information Systems quality

As mentioned above, there is a diversity on IS quality frameworks and models. One recent approach proposes an IS quality framework based on key concepts of IS delivery process: IS delivery paradigm, IS deployment, Software engineering, software production method and systems development methodology [DR06]. IS quality concerns the whole process from the methodology used for IS development, IS development, to the process of transferring a completed system from its development environment into the operational environment. During these phases, several challenges are identified that concern the technical challenge, people challenge and the challenge arisen during the IS delivery cycle. The approach also proposes an IS quality model that describes multidimensional effect of several forces that influence IS quality and IS success.

The most important remark is that the IS quality is impacted by the process management practices employed (i.e. system development methodology, production method and project management) and by people, as competence and experienced IS specialist are central of high-quality IS product [PSV94]. The quality framework is mainly focused by the evaluation of the quality of the process and by the result product at each stage of the life cycle of IS development.

Besides, different quality factors of the other approaches concerning several dimensions of IS quality are also summarized in [DR06] such as timely delivery and relevance beyond deployment, benefits outstrip life-cycle cost, ease of access and use of delivered features, acceptable response times, provision of required functionality and features, reliability of features and high probability of correct and consistent response, maintainability- easily identify sources of defects, etc.

**2.3 Software quality**

The software quality is different from the IS quality. The quality of software emphasises the quality of the production of the artefact whereas the quality of an information system stresses the use of this artefact in an organisational context [VWW93]. The software product quality which is defined by Deming and Juran et al. is "conformance for requirements and fitness for customer use". [As98] defines software quality factors including quality of design, quality of performance and quality of adaptation.

There are a number of models and frameworks for evaluating software quality [Wo06]. In addition, there is a standard ISO 9126 that defines a set of characteristics for software quality evaluation:

- *Functionality* A set of attributes that bear on the existence of a set of functions and their specified properties. The functions are those that satisfy stated or implied needs.

- *Usability* A set of attributes that bear on the effort needed for use, and on the individual assessment of such use, by a stated or implied set of users.

- *Reliability* is the set of attributes that bear on the capability of software to maintain its level of performance under stated conditions for a stated period of time.

- *Efficiency* is the set of attributes that bear on the relationship between the level of performance of the software and the amount of resources used, under stated conditions.

- *Maintainability* is the set of attributes that bear on the effort needed to make specified modifications.

- *Portability* is the set of attributes that bear on the ability of software to be transferred from one environment.

**2.4 Remarks**

Information system quality and software quality are defined with different models and frameworks. Each one includes a set of quality factors. These factors are different from various models, frameworks depending on the perspectives considered. However, what is the rationale of the selection? Why they are chosen? Furthermore, these approaches rarely present impacts among different factors while it helps to manage the quality of IS and to retrieve sources of defects.

# 3 Quality framework in information systems

We developed a quality framework based on the IS architecture. In our framework, the quality factors are selected from the perspectives of user and IS Developer regarding different abstraction levels and views in the IS architecture.

The perspective of user concerns the evaluation of the final product, the requirement satisfaction and the fitness for use. Meanwhile, the perspective of IS developer concerns the quality of different models specified at different abstract levels and the quality of the system implemented.

We illustrate our framework with a very simple example based on the development of an IS for the library management. According to the IS architecture of Zachman, the concerned abstract levels (rows) are requirement analysis, system analysis, system design, implementation and deployment. In this example, we are interested in data and process views (column). An appropriate model is represented in each cell. Along with this architecture, different descriptions at cell level will be presented step by step in followed sections.

**3.1 Quality factors in the framework**

Our framework presents different quality factors concerning various abstract levels according to IS stakeholders' view such as analysis, design, implementation and operation/deployment.

**3.1.1 Quality factor definition**

*a. Requirement quality*

User requirements are analyzed in the early stage of the IS development process. The requirement determination is made by collecting information from conversations with users, collecting existing documents and files or computer-based information [VGH06]. The requirement determination allows to at least understand the issues such as the business objectives; the information needed; when, how and by whom or what data are moved; the rules governing how data are handled and processed [VGH06]. Requirements may be functional, non-functional and interface requirements. Non-qualitative requirements may meet one of the following criteria [Da88]:

-the requirements may be incomplete, inconsistent and/or contain redundancy
-they may not accurately convey the intent of the stakeholders
-in transitioning from the original requirements to the design, the original intention might not be accurately preserved
-over the course of the development of the system, the requirements may change
-the system requirements may not be adequate to meet the needs of the intended application domain.
 -the number and complexity of the set of requirements may tax people's short-term memory beyond its limits
 -the alignment between the requirements for a system, its design, and the implementation may be not preserved.

In our approach, the following criteria of the requirement quality are interesting (defined in [ww1])

- *Completeness*: that means there is enough information to proceed to the next stage or phase of work without risking a serious amount of rework.
- *Consistency*: that means the lack of conflict or contradiction among requirements.
- *Accuracy*: that accurately conveys the intent of the User.

Returning to our example, the user requirements of this system are analyzed as follows:

-Business objectives: management of Reader, Book, Borrowing and Reservation
-The processes description: An available book can be borrowed by readers, if the book is unavailable, the reader can make a reservation. The borrowing time is generally 3 days; however if the reader is staff then the borrowing time is 30 days. The reader can renew her/his borrowing
-The system shall run on the internet, readers can access any time the status of books, consult their borrowing. Only the librarian can make borrowing and reservation.

The requirements captured above are not complete. There is not enough information for the borrowing process. For instance, what states has a book? Is a book blocked for a reservation after it is returned? If it is not blocked, it becomes available then can it be borrowed by others? So how a reservation can be fulfilled? …

Certainly the missing information will impact the succeed stages. Besides, the requirements defined above mean there is no constraint on renewing a borrowing. If this is not the intention of the users then the incompleteness of requirement (i.e. missing constraints on renewing a borrowing) can lead to an inaccuracy of requirements.

*b. Meta-model quality*

In this paper, terminologies *meta-model* and *model* are used according to the OMG four-layer meta-models [OMG02].

A *meta-model* owns constructs and rules that allow specifying or describing a *model*. A model is an instance of its meta-model. In the context of ISA, at the analysis and design levels, different *meta-models* are used to describe different aspects/views of IS such as data view, functional view, organizational view, etc. The result of this description is various *models*. For instance, UML [BRJ98] can be considered as a meta-model that includes several diagrams used to describe the data aspect, functional aspect, process aspect, architectural aspect of an application domain. A class diagram of a concrete application or an activities diagram of a concrete application is a model. Therefore, a meta-model can be considered as a tool that allows the analyst and designer to re-describe user requirements in another form. A "good" meta-model should allows analysts and designers to describe completely, precisely and faithfully what they intend to describe. In the other words, the quality of a meta-model depends on its expressive power and on how it supports designers in the modelling process to obtain a sound specification. Criteria concerning subjective evaluation or social method evaluation (i.e. interview) such as the simplicity and the understand-ability of a meta-model are out of the scope of this paper. Next, we define quality criteria for meta-models as follows:

- *Completeness*: the meta-model allows describing different information that cannot be described (or limited) by other meta-models, this criteria concerns the expressive power of the meta-model. For instance, the class diagram of UML owns constructs allowing describing the object behavior (method), the instantiation relationship and the aggregation relationship. Meanwhile, the Entity-Relationship model [Ch76] (or meta-model) does not; or the class diagram of UML does not describe dynamic specialization and keep track of objects, but the IASDO model does [PDBL06], [PH07].
- *Consistency*: the meta-model is consistent itself and helps designers to obtain a consistent model. Rules defined on a meta-model should ensure the meta-model is consistent and its instances (model) are also consistent. For instance, any process has its input and output information. If the input information does not exist then the output information neither. The IASDO meta-model allows its instance to satisfy this rule.
- *Accuracy*: the meta-model is accurate itself and helps designers to obtain an accurate model. Rules defined on a meta-model should ensure the meta-model is accuracy and its instances (model) are also accuracy. For instance, the Data Flow Diagram has a set of rules [Ce87] to ensure the accuracy of its instance, one of them is "no process can have only input data flows, no process can have only output data flows".

*c. Model quality*

Model quality includes conceptual model quality and logical model quality. Conceptual model quality is defined as "the total quality of features and characteristics of a conceptual model that bears on its ability to satisfy stated or implied needs", or conceptual model quality conforms to requirement established and in how far the requirements of different model-addresses are fulfilled [Mo05]. The logical model is a transformation of the conceptual model into the design level in IS architecture.

In the literature, several works are realized on evaluating the conceptual model quality. Based on this definition and in order to develop a framework for conceptual model quality, [Mo05] synthesized eight different approaches (deductive, codification, inductive, social, analytical, reverse inference, Goal-Question-Metric model and Dromey's methodology). In conclusion they stated that there is no common standard for conceptual model quality.

[MS94], have proposed a quality framework for quality evaluation of data models. This framework is composed of six quality factors: completeness, integrity, understand-ability, simplicity, integration and implement-ability. Each factor is associated a weight, which indicates the importance of the factor. Furthermore, [LSS94] and [KLS95] proposed a quality framework that is focusing on three main criteria: syntax, semantic and pragmatics. Syntactical quality of a model relates to the model-completeness compared to the meta-model, as well as the consistency of the model compared to the meta-model. Semantic quality depends on the relevance of models for the modeling domain (i.e. modeling subject).

There are two characteristics in the semantic quality: *validity* means there are no statements in the model that are not correct and relevant about the domain, and *completeness* means the model does not miss statements that are correct and relevant about the domain. The pragmatic quality of a model depends on the easy comprehension of this model and the feasibility concepts.

In summary, we define the quality criteria of the conceptual model with adaptation from research mentioned above as follows:

- *Completeness*: The completeness of a conceptual model means the model does not miss statements that are correct and relevant about the domain. Certainly, the completeness of a conceptual model depends on the completeness criteria of requirements quality. For instance, the class diagram obtained in the example of library management above will be not complete in relation to the defined requirements if it misses the reader status information (i.e. if a reader is a staff). The activities diagrams are not complete in relation to the user need if these diagrams do not describe *block a book* activity.
- *Accuracy*: The accuracy of a conceptual model means there is no statements in the model that are not correct and relevant about the domain.
- *Consistency*: The model is consistent with the meta-model in the syntax, the model satisfies validation rules of the meta-model (if exist) and there is no contradiction of statements in the model.

These quality factors are evaluated with the application domain (or the user need) rather than with the requirements captured. Returning to our example, if a class diagram of library management system does not describe the *reader status*, then this model is incomplete. If an activity diagram concerning borrowing process does not include the activity *renew*, then this model is incomplete.

*d. Modelling quality*

The modelling quality depends on the experience, skill and objectiveness of modellers. In our framework, the modelling quality can also be understood as the experience and the skill of programmer. [SR98] stated that the subjective position of the modeller is the characterizing issue for the result of the modelling process and proposed a Guidelines of information Modelling including six principals which are Construction adequacy, Economic efficiency, Language adequacy, Systematic design, Clarity and Comparability. The GoM allows the modeller to follow principals in order to reduce the subjectivism in the information modelling process and to improve the quality of information/data models. The result of modelling process is models, specifications or system implemented. Therefore the quality of modelling can be evaluated through the quality of these products.

*e. Data quality*

Data quality is multidimensional within fifteen dimensions [WS96], which can be grouped in four categories: intrinsic data quality, contextual data quality, representational data quality and accessibility data quality. Frequently, data quality is evaluated by widely accepted dimensions such as completeness, accuracy, consistency and timeliness [BP95]; [KLP97].

- *Completeness*: data is not missing and is sufficient breadth and depth for the task at hand. Normally at the database level, the completeness of data is in relation with the structure defined. For example, if data is filled in all attributes of a table then this occurrence is complete. If regarding the domain, the completeness of data is impacted by the completeness criteria of the data (conceptual) model.
- *Consistency*: data is represented in the same format. This criteria also depends on the modelling of data model within integrity rules.
- *Accuracy*: data is correct in its value.
- *Timeliness*: data is sufficiently up-to-date for the task at hand.

The implemented systems produce and transform data. Thus, data quality is considered at the deployment level in IS architecture.

### 3.1.2 Quality factors and different abstract levels

Each defined quality factor in the section above has rationale and is evaluated in an appropriate level in the IS architecture (Fig. 1).

| Quality Factors | **Requirement analysis** | **Analysis** | **Design** | **Implementation** | **Deployment** |
|---|---|---|---|---|---|
| Requirement quality | | | | | |
| Meta-model quality | | | | | |
| Model quality | | | | | |
| Modelling quality | | | | | |
| Data quality | | | | | |

Figure 1. Quality framework in IS based on IS architecture

- **Requirements analysis**: Requirement analysis concerns requirement definition of user. The experience of the analyst impacts the requirement capturing process. So the *requirement quality* and *modelling quality* are considered at this level.
- **Analysis level and design level:** Based on requirements specified, different conceptual models/specifications concerning different IS views are modelled using different meta-models. The user interface, functionalities and logical models are also specified. The experience of analyst and designer impacts the modelling process. The *meta-model quality*, *model quality* and *modelling quality* are considered at these levels.
- **Implementation level:** Based on specifications made in the design level, the IS is developed and implemented. The developer skill impacts the implementation. This level concerns *modelling quality*.
- **Operation/Deployment:** The operating of the system understands the manipulation of data in order to serve user needs. In this level, data quality is taken into account.

## 3.2 Impacts/relations of different quality factors

Figure 2 illustrates the impacts among quality factors and the relation of IS stakeholders' perspective and quality factors in our framework.

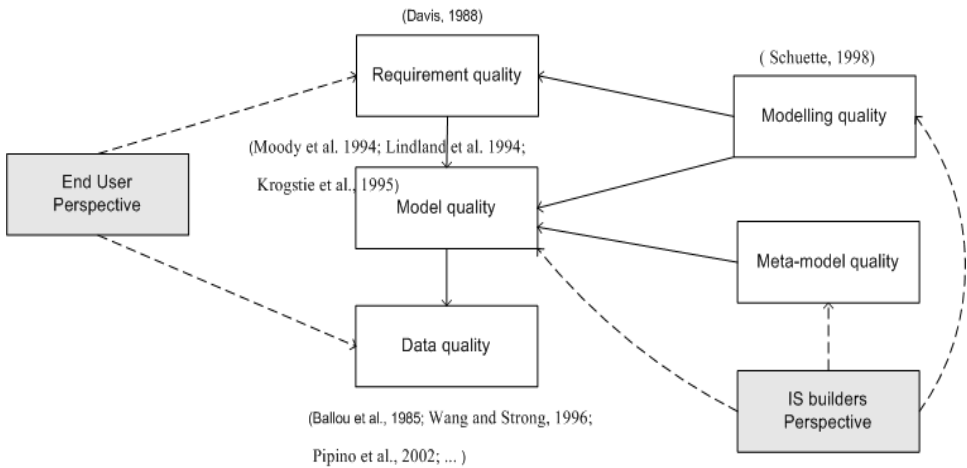

Figure 2. Impacts among quality factors

There are effectively two quality levels of the model quality. The first level is that the obtained models are relevant to described requirements. In other words, these models are complete, accuracy and consistent with defined requirements. The second level concerns the satisfaction of the models to the user needs. If the requirement is not complete, accuracy or consistent with the user need then we cannot obtain qualitative models in relation to the user need even though they can be qualitative in relation to defined requirements. In other words, the requirement quality impacts the model quality.

Data is described by a conceptual data model. The completeness and consistency of data in database may be dependent on the conceptual model. Consequently, the conceptual model may impact the quality of data. For instance, if integrity rules on data consistency are specified with data model and the system implementation respects the specification then the data consistency can be managed. In other words, if the requirements are not well captured/defined, then the system cannot satisfy the users at the end.

The meta-model also impacts the model quality, especially the expressive power property. If the meta-model does not allow describing a certain situation of the application domain then the model can not present that situation. For example, most meta-model for process modelling does not support to describe the responsibility of an organizational role for process execution as well the privilege on data access. So it may occur the inconsistency at the model level such as a role R is responsible for the process execution P that produces information I, but the role R does not have the privilege of creation of I. In this case, the modeller must control it by adding rules. Thus the expressive power of the meta-model impacts on the expressive power of the conceptual model; whether the meta-model includes explicit rules for validating the conceptual model or validating the model implicitly by the experience of its designers. A meta-model of high quality may help to improve or ensure the quality of the conceptual model.

Modelling quality also impacts the model quality. The case above shows that if the rule concerning that situation is not specified by the modeller then the model is inconsistent.

The requirement quality and data quality belong to the end user perspective because the end-user can evaluate them. Meanwhile the model quality, the modelling quality and the meta-model quality belong to Is developer perspective (i.e. analyst, designer, IS programmer, database administrator, etc)

## 4 Conclusion

In this paper, we presented a quality framework of IS, which is based on IS architecture. We define various quality factors in each abstract level that contribute to the IS quality and the relations among them. Our framework shows that the requirement quality is the most important aspect, as it affects the quality of deliverables in the lower levels. With this framework, studying data quality does not only focus on the database level but it also requires focusing on model quality, modelling quality, meta-model quality and requirement quality. This helps to retrieve the root cause of defects. We aim to define a homogeneous set of quality criteria concerning each quality factor. These are widely accepted for information and data quality evaluation such as completeness, consistency and accuracy. However, the meaning is different in each correspondent quality factor. The selection of quality factors in our framework is based on the IS architecture. Indeed, they are concrete and impacted by the user and IS Developer's perspective. Our quality framework allows managing the quality of data and IS.